\documentstyle[preprint,aps]{revtex}

\newcommand{\unit}{\hat{\bf n}}
\newcommand{\pol}{\hat{\bf e}}
\newcommand{\rv}{{\bf r}}
\newcommand{\Ev}{{\bf E}}

\newcommand{\dv}{{\bf d}}

\newcommand{\Dcv}{\hbox{\boldmath$\cal D$}}

\newcommand{\kappav}{\bbox \kappa}

\newcommand{\skv}{{\bf k}}

\newcommand{\kv}{{\bf k}}

\newcommand{\beq}{\begin{equation}}
\newcommand{\eeq}{\end{equation}}
\newcommand{\bea}{\begin{eqnarray}}
\newcommand{\eea}{\end{eqnarray}}

\begin{document}

\draft
\preprint{}
\title{Non-destructive optical measurement of relative phase \\between two
Bose condensates} 
\author{Janne Ruostekoski and Dan F. Walls} 
\address{ Department of Physics,
University of Auckland, Private Bag 92019, \\Auckland, New Zealand}
\date{\today}
\maketitle
\begin{abstract}
We study the interaction of light with two Bose condensates as an open quantum
system. The two overlapping condensates occupy two different Zeeman sublevels and
two driving light beams induce a coherent quantum tunneling between the
condensates. We derive the master equation for the system. It is shown that
stochastic simulations of the  measurements of spontaneously scattered photons
establish the relative phase between two Bose condensates, even though the
condensates are initially in pure number states. These measurements are
non-destructive for the condensates, because only light is scattered, but no
atoms are removed from the system. Due to the macroscopic quantum interference
the detection rate of photons depends substantially on the relative phase
between the condensates. This may provide a way to distinguish,
whether the condensates are initially in number states or in coherent states.

\end{abstract}

\section{Introduction}
In the Bose-Einstein condensation (BEC) phase transition the Bose gas is
expected to acquire nontrivial phase properties. In the conventional reasoning
the condensate is given a macroscopic wave function which acts as a complex order
parameter with an arbitrary but fixed phase \cite{FOR75,LIF80}. The selection of
the phase implicates the spontaneous breakdown of the U(1)
gauge symmetry, breaking the degeneracy of the ground state for a system. The
condensate contains a macroscopic number of particles and the strict conservation
of the number of atoms is abandoned. The removal or addition of one particle in
the condensate is not assumed to affect essentially the system, resulting in
non-vanishing expectation values for particle annihilation and creation operators
in the ground state. These statements are rigorously valid only in the
thermodynamical limit, where the number of particles $N\rightarrow\infty$. 
However, real atom traps have a finite number of atoms, much less than the
classical limit. 

Javanainen and Yoo \cite{JAV96} have shown that interfering two Bose condensates
on an atom detector builds up a quantum coherence between the two condensates
resulting in an interference pattern, even though the condensates are taken to
be in number states with no phases and no violation of the particle number
conservation. This was done by simulating a quantum measurement
process for 1000 atoms. The resulting interference is a consequence of the
correlations between atomic positions and the particular measurement process. It
is necessary that the detected atoms from the two condensates are
indistinguishable, {\it i.e.} one does not know which condensate the detected
atom came from. This uncertainty of the relative atom numbers between the
condensates breaks the global gauge invariance associated with the atom number
conservation. Recently, there has been number of related works on simulating
numerically quantum measurements of the relative phase between two Bose
condensates \cite{NAR96,CIR96,JAC96,WON96,WRI97,YOO97,STE97} (see also
\cite{CAS97}) and photon outputs from single mode cavities \cite{MOL97}. Perhaps,
the most relevant to this paper is the consideration of the Josephson coupling
between the two condensates \cite{JAC96}. Despite the high scientific activity
all the condensate phase simulations have been based on straightforward and rather
idealized models of atom detection. In particular, the measurements have been
{\it destructive}, {\it i.e.} during the measurement process atoms have been
removed from the condensates and the total number of atoms in the two condensates
has not been constant.

Several detection schemes of spontaneously broken gauge symmetry in BEC of
atomic gases have been proposed. Javanainen \cite{JAV96b} has shown that the
amplification of phase coherent laser beams driving Raman transitions between two
Bose condensates is a signature of the broken symmetry. Imamo\=glu and Kennedy
\cite{IMA96} found that coherent spontaneous Raman scattering between two
independent, spatially separated condensates may be eliminated by adjusting the
phase difference of the driving laser beams. In our previous work it was shown
that the relative peak heights in the spectrum of scattered light could possibly
be used to determine the relative condensate phase \cite{RUO97b}, and that the
spontaneous decay rate of independently produced excited atoms depends
strongly on the spontaneously broken symmetry \cite{SAV97}. In all these papers
the condensates were assumed to have macroscopic wave functions with well-defined
phases in accordance with the spontaneous breakdown of gauge symmetry. 

In this paper we consider a {\it gedanken} experiment closely related to the
previously proposed detection schemes of spontaneously broken gauge symmetry in 
BEC \cite{JAV96b,IMA96,RUO97b,SAV97}. We evaluate the conditional probability for
detecting coherent spontaneous scattering of photons from two Bose condensates
driven by external light beams. Spontaneous scattering specifies that the
emission is not stimulated by light, although it is stimulated by atoms
\cite{SAV97}. The conditional probability depends on the macroscopic quantum
coherence and it could possibly be used as a method to detect the phase
difference of the two condensates. However, in our calculations no spontaneously
broken gauge symmetry is initially assumed, but the relative phase of the two
Bose condensates is established via direct measurements of spontaneously
scattered photons. This measurement technique is {\it non-destructive} for the
condensates; the light scatters atoms between the condensates, but the total
number of atoms in the two condensates is conserved. Even though we start
initially from number states without any phase information and without the
violation of the particle number conservation, the condensates behave as if they
had a phase. A non-destructive optical detection of a Bose condensate has been
reported recently \cite{AND96}. Coherent forward scattering was measured by
blocking the transmitted probe beam by a thin wire. Our scheme also relates
closely to the experimental production of two overlapping condensates in the same
trap using nearly lossless sympathetic cooling of one state via thermal contact
with the other evaporatively cooled state \cite{MYA96}. The two species were the 
$|F=1,m=-1\rangle$ and $|F=2,m=2 \rangle$ hyperfine spin states of $^{87}$Rb.
Similarly to Refs.~{\cite{JAV96b,RUO97b}} we assume that the atoms are confined
in the same trap and that they occupy two different Zeeman sublevels which are
optically coupled through a common excited state by two low intensity
off-resonant light beams.  

In the recent measurements Andrews {\it et al.} \cite{AND97} have found 
evidence of macroscopic quantum coherence in BEC by interfering two
condensates. The two condensates had the same internal atomic states, but
different external wave functions. The spatially separated condensates were
created by focusing a blue  detuned laser into the center of the magnetic trap,
generating a repulsive optical force. The interference fringes were measured
optically by short resonant laser pulses (absorption imaging), after the
condensates had expanded ballistically about 40 ms. If the condensates were
initially in number states, the observed interference was a consequence of the
uncertainty in which condensate the atoms absorbed photons. Similarily, in our
scheme the macroscopic quantum interference results from the uncertainty in the
initial state of atoms, when they are excited by the driving light beams.

We begin in Sec.~{\ref{effham}} by introducing the Hamiltonian for the system.
In the limit of large detunings of the driving light fields from the atomic
resonances the excited state operators may be eliminated adiabatically. We
obtain an effective two-state Hamiltonian between the two Bose condensates. The
dynamics of the two condensates and the driving light is considered as an open
quantum system in Sec.~{\ref{mas}}. The many-particle master equation is derived
by eliminating the vacuum modes of the electromagnetic fields. In
Sec.~{\ref{josep}} we discuss the light-induced coherent quantum tunneling
between the Bose condensates. In Sec.~{\ref{stoch}} the evolution of the master
equation in terms of stochastic trajectories of state vectors is studied. We show
that the conditional probability for the detection of the $(n+1)^{\rm th}$ photon
during the time interval $[t,t+\delta t]$, given that the $n^{\rm th}$ detection
occured at the time $t_{n}$ depends strongly on the relative phase between the
condensates. The results of simulations are presented in Sec.~{\ref{res}}. In the
simulations the relative phase between the two Bose condensates is established by
the measurements of spontaneously scattered photons, even though the condensates
are initially in pure number states. We also propose a test which could possibly
distinguish an initial number state from an initial coherent state, {\it i.e.} to
test the validity of the conventional symmetry breaking arguments in BEC of
dilute atomic gases. Finally, a few concluding remarks are made in
Sec.~{\ref{conc}}.

\section{Effective two-level Hamiltonian}
\label{effham}

We consider two spatially overlapping Bose condensates, whose ground states differ
in their internal quantum numbers. We assume the condensates in two
different Zeeman sublevels  $|b\rangle=|g,m\rangle$ and $|c\rangle=|g,m-2\rangle$
\cite{JAV96b,RUO97b}, see Fig.~{\ref{levelfig}}. The state $c$ is optically
coupled to the electronically excited state $|e\rangle=|e,m-1\rangle$ by the
driving field $\Ev_{d2}$ having a polarization $\sigma_+$ and a dominant
frequency $\Omega_2$. Similarly, the state $b$ is coupled to $e$ by the driving
field $\Ev_{d1}$ with a polarization $\sigma_-$ and a dominant frequency
$\Omega_1$. We assume that the phase-coherent light fields $\Ev_{d1}$ and
$\Ev_{d2}$ propagate in the positive $z$ direction and are detuned far from the
resonances of the corresponding atomic transitions. The light fields are also
assumed to be in  coherent states. We only consider the coherent spontaneous
scattering between the condensates, which is stimulated by a large number of
atoms in the condensates. The decay into the non-condensate center-of-mass (c.m.)
states is also stimulated by  the Bose-Einstein statistics. However, at very low
temperatures this stimulation is much weaker, because most of the particles are
in the condensates. The Bose stimulated spontaneous emission into the
non-condensate modes also scatters  photons into the solid angle of $4\pi$,
whereas in the decay into the condensates the photons are emitted into a narrow
cone in the forward direction. In addition to the Bose stimulation of the
spontaneous emission there is the unstimulated free space decay $\gamma$, which
is always present. This also scatters photons into the solid angle of $4\pi$, and
with a sufficiently large number of atoms in  the two condensates the free space
decay may be ignored. The Hamiltonian for the system reads
\cite{RUO97b,JAV95b,RUO97a}    
\bea  
H &=& \hbar\omega_{cb}\,
c^\dagger c +\sum_{\skv}\hbar(\omega_{eb}+\epsilon_{\skv})\, e_{\skv}^\dagger
e_{\skv} +\sum_q\hbar\omega_q\, a^\dagger_q a_q\nonumber\\
&-&\sum_{\skv}\left(\int d^3r\,\dv_{be}\cdot\Ev_1(\rv)
\phi^*_b(\rv)\phi_{e\skv}(\rv) \,b^\dagger e_{\skv}+ {\rm H.c.}\right)\nonumber\\
&-&\sum_{\skv}\left(\int d^3r\,\dv_{ce}\cdot\Ev_2(\rv)
\phi^*_c(\rv)\phi_{e\skv}(\rv)\, c^\dagger e_{\skv}+{\rm H.c.}\right)\,,
\label{eq:HDN}  
\eea 
where the creation operators for the atomic states $e$, $b$,
and $c$ are given by $e^\dagger$, $b^\dagger$, and $c^\dagger$, respectively. The
condensate wave functions, which would typically be solutions of the
Gross-Pitaevskii equation \cite{LIF80}, are $\phi_b$ and $\phi_c$. The excited
state wave function for the c.m. state $\kv$ is $\phi_{e\skv}$ with the c.m.
energy $\epsilon_{\skv}$. The photon annihilation operator for the mode $q$ is
$a_q$. The internal energies are described by the frequencies of the optical
transitions  $e\rightarrow b$ and $e\rightarrow c$, which are $\omega_{eb}$ and
$\omega_{ec}$ ($\omega_{cb}=\omega_{eb}-\omega_{ec}$). The last two terms in 
Eq.~{(\ref{eq:HDN})} are for the atom-light dipole interaction. The dipole matrix
element for the atomic transition $e\rightarrow b$ is given by $\dv_{be}$.

In the limit of large detunings, $\Delta_1=\Omega_1-\omega_{eb}$ and
$\Delta_2=\Omega_2-\omega_{ec}$, the excited state operators $e_{\skv}$ in
Eq.~{(\ref{eq:HDN})} may be eliminated adiabatically, and the c.m.
energies of the excited state may be ignored \cite{JAV95b}. We insert the
steady-state solutions of the Heisenberg equations of motion for the
slowly varying operators $\tilde{e}_{\skv}\equiv e^{i\Omega_1 t} e_{\skv}$ into
the Hamiltonian {(\ref{eq:HDN})} as in Refs.~{\cite{JAV96b,RUO97b}}. With the
help of the completeness of the c.m. states $\phi_{e\skv}$,
Eq.~{(\ref{eq:HDN})} reduces to an effective two-state Hamiltonian 
\bea 
H &=& \hbar\omega_{cb}\, c^\dagger c
+\sum_q\hbar\omega_q\, a^\dagger_q a_q-{1\over\hbar\Delta_1}\left\{b^\dagger b
\int d^3r\,\dv_{be}\cdot{\bf E}_1(\rv)\,\dv_{eb}\cdot{\bf E}_1
(\rv)\phi^*_b(\rv)\phi_b(\rv)\right.\nonumber\\ &+& c^\dagger
c \int d^3r\,\dv_{ce}\cdot{\bf E}_2(\rv)\,\dv_{ec}\cdot{\bf
E}_2(\rv)\phi^*_c(\rv) \phi_c(\rv) \nonumber\\ &+&\left.\left(b^\dagger
c\int d^3r\,\dv_{be}\cdot{\bf E}_1(\rv)\,\dv_{ec}\cdot{\bf
E}_2(\rv)\phi^*_b(\rv) \phi_c(\rv)+{\rm H.c.}\right)\right\}\,.
\label{eq:HDN2}  
\eea 

The electric fields may be solved from Eq.~{(\ref{eq:HDN2})} by integrating the
Heisenberg equations of motions \cite{JAV95b}. It is advantageous first to
postpone the rotating wave approximation (RWA) and then to {\it define} the
positive frequency parts of the light fields by the dominant time dependence of
the operators. The positive frequency component of the total electric field ${\bf
E}^+_{1}= {\bf E}^+_{d1}+{\bf E}^+_{s1}$ is expressed in terms of the classical
driving field ${\bf E}^+_{d1}$, that would prevail in the absence of matter, and
the scattered field ${\bf E}^+_{s1}$. The radiation reaction effects \cite{RUO97a}
may  be ignored in the limit of large detuning, when the multiple scattering is
negligible. The driving field inside the sample in the {\it length} gauge should
be understood as the driving electric displacement divided by the permittivity of
the vacuum \cite{RUO97a,RUO97c}. Assuming the light is measured outside the
sample the scattered fields are given by    
\begin{mathletters}  
\beq  
\tilde{\bf
E}^+_{s1}({\bf r}) = {1\over\hbar\Delta_1}\int d^3r'\, {\bf K}({\bf d}_{be};{\bf
r}-{\bf r'})\phi^*_b({\bf r'})\left\{ \phi_b({\bf
r'})\dv_{eb}\cdot\tilde{\bf E}^+_{d1}(\rv')\,b^\dagger b +\phi_c({\bf
r'})\dv_{ec}\cdot\tilde{\bf E}^+_{d2}(\rv') \,b^\dagger\tilde{c}\right\}\,,
\label{eq:FEQa}
\eeq
\beq
\tilde{\bf E}^+_{s2}({\bf r}) = {1\over\hbar\Delta_1}\int d^3r'\,
{\bf K}({\bf d}_{ce};{\bf r}-{\bf r'})\phi^*_c({\bf r'})\left\{\phi_c({\bf
r'})\dv_{ec}\cdot \tilde{\bf E}^+_{d2}(\rv')\,\tilde{c}^\dagger \tilde{c}
+\phi_b({\bf r'})\dv_{eb}\cdot\tilde{\bf E}^+_{d1}(\rv')
\,\tilde{c}^\dagger b\right\}\,,
\label{eq:FEQb}
\eeq
\label{eq:FEQ}
\end{mathletters}
where we have defined the slowly
varying operators $\tilde{\bf E}^+_1 \equiv e^{i\Omega_1 t}{\bf E}^+_1$,
$\tilde{\bf E}^+_2 \equiv e^{i\Omega_2 t}{\bf E}^+_2$, and
$\tilde{c} \equiv e^{i(\Omega_1-\Omega_2) t}c$.
We have made the first Born approximation and replaced the electric fields under
the integrals in Eq.~{(\ref{eq:FEQ})} by the corresponding driving fields.
The kernel ${\bf K}(\Dcv;{\bf r})$ coincides with the classical
expression \cite{JAC75} of the positive-frequency component of the electric 
field from a monochromatic dipole with the complex amplitude $\Dcv$, given 
that the dipole resides at the origin and the field is observed at $\rv\ne0$. The
explicit expression is 
\begin{equation} 
{\bf K}(\mbox{\boldmath$\cal D$};{\bf r}) =
{1\over4\pi\epsilon_0}
\bigl\{ k^2(\hat{\bf n}\!\times\!\mbox{\boldmath$\cal
D$})\!\times\!\hat{\bf n}{e^{ikr}\over r} +[3\hat{\bf n}(\hat{\bf
n}\cdot\mbox{\boldmath$\cal D$})-\mbox{\boldmath$\cal D$}]
\bigl( {1\over r^3} - {ik\over r^2}\bigr) e^{ikr}
\bigr\},
\label{eq:DOL}
\end{equation}
where $k=\Omega /c$ and $\unit$ is a unit vector pointing from the source point
toward the field point. 

\section{Master equation}
\label{mas}

In this section we consider the dynamics of the two Bose condensates and the
driving light fields as an open quantum system and eliminate the vacuum modes.
We assume the driving electric fields to be plane waves   
\beq
\tilde{\bf E}^+_{d1}({\bf r}) =\hbox{$1\over2$}
{\cal E}_1 \pol_- 
e^{i {\bbox \kappa}_{1}
\cdot {\bf  r}},\quad
\tilde{\bf E}^+_{d2}({\bf r}) =\hbox{$1\over2$}
{\cal E}_2 \pol_+ 
e^{i {\bbox \kappa}_{2}
\cdot {\bf r}}\,.
\label{eq:INF}
\eeq
We insert ${\bf E}^+_{1}={\bf E}^+_{d1}+{\bf E}^+_{s1}$ and ${\bf
E}^+_{2}= {\bf E}^+_{d2}+{\bf E}^+_{s2}$ into the Hamiltonian
(\ref{eq:HDN2}), where ${\bf E}^+_{s1}$ and ${\bf E}^+_{s2}$ are now
considered as vacuum fields, and keep only the terms of second order in
$1/\Delta_1$, {\it i.e.} we neglect the terms, which contain a product of two
scattered fields. The system Hamiltonian $H_S$ for the condensates and the
macroscopic light fields is coupled to the reservoir of the vacuum fields by the
Hamiltonian $H_{SR}$. The total Hamiltonian is given by $H=H_S+H_R+H_{SR}$,
where the different parts have the following expressions 
\bea
&&H_{SR}= -{1\over\hbar\Delta_1}\left\{
\int d^3r\,\dv_{eb}\cdot{\bf E}^+_{s1} (\rv)\left(\dv_{be}\cdot{\bf
E}^-_{d1}(\rv)\,\phi^*_b(\rv)\phi_b(\rv)\,b^\dagger b+\dv_{ce}\cdot
{\bf E}^-_{d2}(\rv)\phi^*_c(\rv) \phi_b(\rv)\,{c}^\dagger
b\right)+{\rm H.c.}\right\}\nonumber\\ &&\,\,\,\,
-{1\over\hbar\Delta_1}\left\{ \int d^3r\,\dv_{ec}\cdot{\bf
E}^+_{s2}(\rv)\left(\dv_{ce}\cdot{\bf E}^-_{d2}(\rv)\,\phi^*_c(\rv)
\phi_c(\rv)\,{c}^\dagger {c}+\dv_{be}\cdot{\bf E}^-_{d1}(\rv)
\,\phi^*_b(\rv) \phi_c(\rv)\,b^\dagger {c}\right)+{\rm
H.c.}\right\}\,, \nonumber\\   &&H_S= \hbar(\omega_{cb}-\delta_2)\, 
{c}^\dagger {c}-\hbar\delta_1 b^\dagger b +\left(\hbar\kappa\,
b^\dagger {c}+{\rm H.c.} \right),\quad H_R=\sum_q\hbar\omega_q\, a^\dagger_q
a_q\,.   
\label{eq:HDN3}   
\eea 
Here we have used the following notation   
\beq
\delta_1={|{\cal E}_1|^2d_{eb}^2\over 4\hbar^2\Delta_1},\quad 
\delta_2={|{\cal E}_2|^2d_{ec}^2\over 4\hbar^2\Delta_1},\quad 
\kappa={{\cal E}^*_1{\cal E}_2d_{eb}d_{ec}\over 4\hbar^2\Delta_1}\int d^3 r\, 
\phi^*_b (\rv)e^{-i {\bbox \kappa}_{12}
\cdot{\bf r}}\phi_c(\rv)\,,
\label{eq:para}
\eeq
where $\kappav_{12}=\kappav_1-\kappav_2$ is the wavevector difference of the
incoming light fields, and $\delta_1$ and $\delta_2$ are the light-induced level
shifts. The dipole matrix element $d_{eb}$ contains the reduced dipole matrix
element and the corresponding nonvanishing Clebsch-Gordan coefficient. 

We assume a reservoir at $T=0$ and obtain the equation of motion for the reduced
density matrix of the system in the interaction picture $\tilde{\rho}_S$ 
using the standard Born and Markov approximations 
\beq
\dot{\tilde{\rho}}_S=-{1\over\hbar^2}\int^\infty_0 d\tau\, {\rm Tr}_R\left\{
[\tilde{H}_{SR}(t),[\tilde{H}_{SR}(t-\tau),\tilde{\rho}_{S}(t)
\otimes\rho_{R}]\,]\right\}\,, 
\label{eq:mas}
\eeq 
where the trace is calculated over the vacuum modes and the reduced density matrix
of the reservoir $\rho_R$ is assumed to be in a statistical mixture of
eigenstates of $H_R$ \cite{COH92}. The calculation of Eq.~{(\ref{eq:mas})} reduces
to the evaluation of the vacuum field correlations
\beq   
P(r_{12},\Omega_1,\dv_{eb})\equiv {1\over\hbar^2}\int^\infty_0 d\tau\,
e^{i\Omega_1\tau}\langle \dv_{eb}\cdot{\bf E}^+_{s1} (\rv_1,t)\,\dv_{be}\cdot{\bf
E}^-_{s1} (\rv_2,t-\tau)\rangle\,.
\label{eq:prevac}
\eeq
Lenz {\it et al.} {\cite{LEN94}} have derived the master equation for light matter
interactions in the RWA. Applying the RWA leads to spatially non-local Cauchy
principal value terms, which also come out in the integration of the scattered
electric fields  {(\ref{eq:DOL})}, in addition to the dipole radiation kernel, if
the straightforward RWA is performed \cite{JAV95b}. To avoid obvious physical and
mathematical problems of the RWA we include in the vacuum electromagnetic fields
the negative mode frequencies ($\omega_q<0$) for the photon annihilation and
creation operators. This is basically equivalent to ignoring some of the
commutation relations between photon annihilation and creation operators. 

We assume that there is a cutoff in the wave numbers $q$ of the photons; we
multiply the density of the states of the electromagnetic fields by
$e^{-q^2\alpha^2/4}$, with
$\alpha > 0$ being a length scale. The cutoff removes all mathematical
problems concerning, e.g., the exchange of the order of derivatives
and integrals, which are abundant in the theory without the cutoff. At
the end of the calculations we ultimately take the limit
$\alpha\rightarrow0$. 

The evaluation of Eq.~{(\ref{eq:prevac})} is very similiar to the integration of
the electric field in Ref.~{\cite{JAV95b,RUO97a}}. The time evolution of the
vacuum fields in the interaction picture is defined by $H_R$. First, we set
$r_{12}\equiv |\rv_1-\rv_2|=0$. For a small but nonzero $\alpha$, the result is
\beq
P(r_{12}=0,\Omega_1,\dv_{eb})={d_{eb}^2\omega_0^3\over6\pi\hbar\epsilon_0 c^3}
+  i{2 d_{eb}^2\sqrt\pi\over3\pi^2 \epsilon_0\hbar\alpha^3}\equiv \gamma_{eb}+i\,
\delta\omega_{eb}
\label{eq:local}
\eeq
The first term $\gamma_{eb}$ in Eq.~{(\ref{eq:local})} is the familiar
spontaneous linewidth of the atomic transition $e\rightarrow b$.
The second term $\delta\omega_{eb}$ diverges as the photon momentum cutoff goes
to infinity with $\alpha\rightarrow0$. This part, after a proper renormalization,
contributes to the Lamb shift. From now on we assume that the Lamb shifts are
already included in the transition frequencies, and ignore the $\delta\omega_{eb}$
term in Eq.~{(\ref{eq:local})}.

The expression {(\ref{eq:prevac})} is not divergent for $r_{12}\neq0$. In that
case we obtain    
\beq 
P(r_{12},\Omega_1,\dv_{eb})= -{ic\over
4\pi\epsilon_0\hbar}\,\dv_{eb}\cdot(\dv_{be}
\times\mbox{\boldmath$\nabla$})\times\mbox{\boldmath$\nabla$}\, {
e^{ik_1 r_{12}}\over r_{12}} = -{i\over \hbar}\,
\dv_{eb}\cdot{\bf K}(\dv_{be};r_{12})\,,
\label{eq:vaccor}     
\eeq
The value of {\it not} having made the RWA in the evaluation of
Eq.~{(\ref{eq:prevac})} emerges in the apparent physical results
Eqs.~{(\ref{eq:local})} and {(\ref{eq:vaccor})}. 

By defining the operators
\begin{mathletters}
\bea
C_1(\rv) &\equiv & -{1\over\hbar\Delta_1}\left(\dv_{eb}\cdot\tilde{\bf
E}^+_{d1}(\rv)\,\phi^*_b(\rv)\phi_b(\rv)\,b^\dagger b+\dv_{ec}\cdot
\tilde{\bf E}^+_{d2}(\rv)\phi^*_b(\rv)\phi_c(\rv) \, b^\dagger \tilde{c}
\right)\,,
\\ C_2(\rv) &\equiv &  -{1\over\hbar\Delta_1}\left(\dv_{ec}\cdot\tilde{\bf
E}^+_{d2}(\rv)\, \phi^*_c(\rv) \phi_c(\rv)\,\tilde{c}^\dagger
\tilde{c}+\dv_{eb}\cdot\tilde{\bf E}^+_{d1}(\rv) \,\phi^*_c(\rv)
\phi_b(\rv)\,\tilde{c}^\dagger b\right)\,, 
\eea
\label{eq:c1c2}
\end{mathletters}
the equation of motion for the reduced density matrix is obtained from 
Eqs.~{(\ref{eq:mas})}, {(\ref{eq:local})}, and~{(\ref{eq:vaccor})} 
\bea
\dot{\rho}_S &=&
-{i\over\hbar}[\tilde{H}_S,\rho_S]\nonumber\\&&\mbox{}+{i\over\hbar}\int d^3
r_1\,d^3 r_2\,\dv_{eb}\cdot{\bf K}(\dv_{eb}; r_{12})  \left(
C_1^\dagger(\rv_1) C_1(\rv_2)\rho_S- C_1(\rv_2)\rho_S C_1^\dagger(\rv_1)\right)
\nonumber\\ &&\mbox{}-{i\over\hbar}\int  d^3 r_1\,d^3 r_2\,\dv_{eb}\cdot{\bf
K}^*(\dv_{eb};  r_{12})   \left(\rho_S C_1^\dagger(\rv_1) C_1(\rv_2)-
C_1(\rv_2)\rho_S C_1^\dagger(\rv_1) \right)\nonumber\\  
&&\mbox{}+{i\over\hbar}\int d^3 r_1\,d^3 r_2\,\dv_{ec}\cdot{\bf K}(\dv_{ec};
r_{12})  \left( C_2^\dagger(\rv_1) C_2(\rv_2)\rho_S- C_2(\rv_2)\rho_S
C_2^\dagger(\rv_1)\right) \nonumber\\ &&\mbox{}-{i\over\hbar}\int  d^3 r_1\,d^3
r_2\,\dv_{ec}\cdot{\bf K}^*(\dv_{ec};  r_{12})   \left(\rho_S
C_2^\dagger(\rv_1) C_2(\rv_2)- C_2(\rv_2)\rho_S C_2^\dagger(\rv_1) \right)
\label{eq:mas2}  
\eea
where the terms containing ${\bf K}(\dv; r_{12})$ should
be understood in such a way that the divergent in-phase part of the dipole
field at $r_{12}=0$ is ignored. In the limit  $r_{12}\rightarrow0$ the real part
of $i{\bf K}(\dv_{eb}; r_{12})/\hbar$ coincides with $\gamma_{eb}$, the real
part of  Eq.~{(\ref{eq:local})}. The Hamiltonian $\tilde{H}_S$ is the system
Hamiltonian Eq.~{(\ref{eq:HDN3})} in terms of the slowly varying operators
$\tilde{c}$ \beq
\tilde{H}_S= -\hbar(\delta_{cb}+\delta_2)\, 
{c}^\dagger {c}-\hbar\delta_1 b^\dagger b +\left(\hbar\kappa\,
b^\dagger {c}+{\rm H.c.} \right)\,,
\label{eq:nsys}
\eeq
where we have defined the two-photon detuning $\delta_{cb}= 
\Omega_1-\Omega_2-\omega_{cb}$.

The non-local interaction terms in Eq.~{(\ref{eq:mas2})} depending on  ${\bf
K}(\dv; r_{12})$ represent the dipole-dipole interactions between atoms. According
to Ref.~{\cite{JAV95b}}, if the detunings $\Delta_1$ and $\Delta_2$ are at least
comparable to the collective linewidth of the Bose condensate $\Gamma$, and if
the characteristic size of the condensate $l$ is substantially larger than the
wavelength of light $l\gg \lambda$, the dipole-dipole interactions and the
multiple scattering of light may be ignored. The condensate collective
linewidth has been estimated to be $\Gamma=3N \gamma /(2 l^2k^{2})$, where
$\gamma$ is the transition's free space natural linewidth and $N$ is the total
number of atoms \cite{JAV94,YOU95}. We assume these conditions to be valid and
neglect the dipole-dipole interactions between the atoms. Because in the limit of
low light intensity the contact  interactions are inconsequential in the
interactions between dipole atoms \cite{RUO97c}, this is done by ignoring the
terms depending on  ${\bf K}(\dv; r_{12})$ with $r_{12}\neq0$ in
Eq.~{(\ref{eq:mas2})}. We obtain for the master equation  \bea 
\dot{\rho}_S &=& -{i\over\hbar}[\tilde{H}_S,\rho_S]
-\gamma_{eb}\int d^3 r \,\left(
C^\dagger_1(\rv)C_1(\rv)\rho_S +\rho_S C^\dagger_1(\rv)C_1(\rv) -2C_1(\rv) \rho_S
C^\dagger_1(\rv) \right)\nonumber\\&&\mbox{}-\gamma_{ec}\int d^3 r \,\left(
C^\dagger_2(\rv)C_2(\rv)\rho_S +\rho_S C^\dagger_2(\rv)C_2(\rv) -2C_2(\rv) \rho_S
C^\dagger_2(\rv) \right)\,.   
\label{eq:mas3}  
\eea

\section{Josephson effect}
\label{josep}

In the absence of the coupling to the reservoir the dynamics of the system
determined by the system Hamiltonian $H_S$ in Eq.~{(\ref{eq:HDN3})} can be solved
analytically. The solutions for the ground state annihilation operators
$\tilde{c}_{\skv}$ and  $b_{\skv}$ are
\begin{mathletters}
\beq
\tilde{c}(t)=e^{i(\bar{\delta}+\delta_1)t}
\left\{\tilde{c}(0)\left(\cos{\Omega_R t}
+{i\bar{\delta}\over \Omega_R}\sin{\Omega_R t}\right)-{i\kappa\over\Omega_R}
b(0)\sin{\Omega_R t}\right\}\,,
\eeq
\beq
b(t)=e^{i(\bar{\delta}+\delta_1)t}
\left\{b(0)\left(\cos{\Omega_R t}
-{i\bar{\delta}\over \Omega_R}\sin{\Omega_R t}\right)-{i\kappa\over\Omega_R}
\tilde{c}(0)\sin{\Omega_R t}\right\}\,,
\eeq
\label{eq:osc}
\end{mathletters}
where we have defined "one-half of the effective two-photon detuning"
$\bar{\delta}=(\delta_{cb}-\delta_1+\delta_2)/2$. The oscillation frequency in 
Eq.~{(\ref{eq:osc})} is given by $\Omega_R=(\bar{\delta}^2+\kappa^2)^{1/2}$. To 
simplify the algebra, we have assumed $\kappa$ to be real.

Before the light is switched on, the atoms in the states $b$ and $c$ are 
assumed to be uncorrelated. The driving light fields induce a coupling between the
two Bose condensates, which is analogous to the coherent tunneling of Cooper pairs
in a Josephson junction \cite{JAV86}. According to the Josephson effect, the 
atom numbers of the condensates oscillate even if the number of atoms in each
condensate is initially equal. This can be seen easily by using the basic
spontaneous symmetry breaking arguments and assuming the condensates to be in 
coherent states. Then, the annihilation operators have nonvanishing expectation
values at the time $t=0$: 
\beq
\langle b(0)\rangle=\sqrt{N_b}\,e^{i\varphi_b},\quad
\langle c(0)\rangle= \sqrt{N_c}\,e^{i\varphi_c}\,.
\label{eq:SSB}
\eeq 
We have assumed that the
condensates in the states $b$ and $c$ have the expectation values for the number
operators $N_b$ and $N_c$, respectively. Now, if $N_b=N_c=N/2$, where $N$ is the
total number of atoms, the expectation value for the occupation number in the
condensate $c$ is obtained from Eq.~{(\ref{eq:osc})} 
\beq
\langle \tilde{c}^\dagger(t)\tilde{c}(t) \rangle ={N\over 2} \left( 1-
{\kappa\over\Omega_R}\sin{(\varphi)}\sin{(2\Omega_Rt)}-
{2\kappa\bar{\delta}\over\Omega^2_R}\cos{(\varphi)}\sin^2{(\Omega_Rt)}
\right)\,,
\label{eq:jose}
\eeq
where $\varphi\equiv\varphi_c-\varphi_b$. The oscillation in 
Eq.~{(\ref{eq:jose})} is a consequence of the macroscopic quantum coherence. In
each measurement process the relative phase between the two condensates is
selected as a random number. The value of the phase difference determines the
amplitude of the oscillations according to Eq.~{(\ref{eq:jose})}. 

\section{Stochastic Schr\"odinger equation}
\label{stoch}

In this section we study the evolution of the master equation (\ref{eq:mas3}) in
terms of stochastic trajectories of state vectors \cite{DAL92,GAR92,CAR93}. The
dissipation of energy from the quantum system of macroscopic light fields and the
two Bose condensates is described by the coupling to a zero temperature reservoir
of vacuum modes, resulting in the spontaneous linewidth for atoms. The master
equation {(\ref{eq:mas3})} is in the Lindblad form \cite{LIN76} and is equivalent
to the Monte-Carlo evolution of wave functions (MCWF) \cite{DAL92}. The MCWF
procedure consists of the evolution of the system with a non-hermitian
Hamiltonian $H_{\rm eff}$, and randomly decided quantum 'jumps', followed by wave
function normalization. In our case the quantum jumps correspond to the
detections of spontaneously emitted photons. The system evolution is thus
conditioned on the outcome of a measurement. The non-hermitian Hamiltonian is
obtained from  Eq.~{(\ref{eq:mas3})}  
\beq 
H_{\rm eff}=\tilde{H}_S-i\hbar \left(\gamma_{eb}\int d^3 r\,
C^\dagger_1(\rv)C_1(\rv)+ \gamma_{ec}\int d^3 r\, 
C^\dagger_2(\rv)C_2(\rv)\right)\,. \label{eq:nonher}
\eeq
The non-unitary evolution corresponds to the modification of the state of the
system associated with a zero detection result of the spontaneously emitted
photons. Because the output is being continuously monitored by the detector, we
gain information about the system even if no photons have been emitted.

The operators $C_1$ and  $C_2$, from Eq.~{(\ref{eq:c1c2})}, correspond to the
excitations of atoms by the driving light beams followed by the emissions of the
photons with the polarizations $\sigma_-$ and $\sigma_+$, respectively. In the
translationally invariant system and for a non-interacting gas the spontaneous
emission, Eq.~{(\ref{eq:FEQ})}, is directed exactly parallel to the positive
$z$ axis, because of the momentum conservation. In that direction the two
polarizations are perfectly distinguishable. In a finite size trap the
uncertainty in momenta introduces a narrow scattering cone for spontaneously
emitted photons. For simplicity, we assume that this cone is narrow enough, so
that the two polarizations can still be approximated as perfectly
distinguishable. We also assume that all the spontaneously emitted photons can be
detected. Then, we have two detection channels corresponding to the two different
polarizations.

The state vector at the time $t$ is denoted by $\psi_{\rm sys}(t)$. Its evolution
is determined by the non-hermitian Hamiltonian $H_{\rm eff}$, defined in
Eq.~{(\ref{eq:nonher})}. If the wave function $\psi_{\rm sys}(t)$ is normalized,
the probability that a photon with the polarization $\sigma_-$ is detected
during the time interval $[t,t+\delta t]$ is  
\beq 
P_-=2\gamma_{eb}\int d^3 r\,\langle
\psi_{\rm sys}(t)\, |\, C^\dagger_1(\rv)C_1(\rv)\, |\, \psi_{\rm sys}(t)\rangle\,
\delta t\,. 
\label{eq:prob1}
\eeq 
Similarly, the probability for the detection of a photon with
the polarization $\sigma_+$ is 
\beq 
P_+=2\gamma_{ec}\int d^3 r\,\langle \psi_{\rm
sys}(t)\,| \,C^\dagger_2(\rv)C_2(\rv)\,|\, \psi_{\rm sys}(t)\rangle\, \delta t\,.
\label{eq:prob2}
\eeq
The probability of no detections is $1-P_--P_+$.

In the implementation of the simulation algorithm we first, at the time $t_0$,
generate a quasi-random number $\epsilon$ which is uniformly distributed between
0 and 1. We assume that the state vector $\psi_{\rm sys}(t_0)$ at the time $t_0$
is normalized. Then, we evolve the state vector by $H_{\rm eff}$
iteratively for finite time steps $\Delta t\simeq\delta t$. At each time step
$n$ we compare $\epsilon$ to the reduced norm  
\beq
\langle \psi_{\rm sys}(t_0+n\Delta t)\, |\, 
\psi_{\rm sys}(t_0+n\Delta t)\rangle=
\langle \psi_{\rm sys}(t_0)\, |\,  
e^{iH_{\rm eff}^\dagger n\Delta t/\hbar }e^{-i H_{\rm eff} n\Delta t/\hbar
}\,|\,  \psi_{\rm sys}(t_0)\rangle\,,
\label{eq:norm}
\eeq  
until $\langle \psi_{\rm sys}(t_0+t)\, |\, \psi_{\rm sys}(t_0+t)\rangle
<\epsilon$, when the detection of a photon occurs. It is easy to show that, at
first order in $\Delta t$, Eq.~{(\ref{eq:norm})} is the joint probability of
not detecting photons in any of the $n$ time intervals. After the detection we
generate a new quasi-random number $\eta$. We evaluate $P_-$ from
Eq.~{(\ref{eq:prob1})} and $P_+$ from Eq.~{(\ref{eq:prob2})} at the time of the
detection. If $\eta<P_-/(P_-+P_+)$, we say the polarization of the detected
photon is $\sigma_-$. If the photon has been observed during the time step
$t\rightarrow t+\Delta t$, we take the new wave function at $t+\Delta t$:    \beq 
|\, \psi_{\rm
sys}(t+\Delta t)\rangle=\sqrt{2\gamma_{eb}}\int d^3 r\,C_1(\rv)\,|\, \psi_{\rm
sys}(t)\rangle\,, 
\label{eq:meas1}
\eeq
which is then normalized. Otherwise, $\eta>P_-/(P_-+P_+)$ and the polarization
of the detected photon is $\sigma_+$. In that case the new wave function before
the normalization reads 
\beq 
|\, \psi_{\rm sys}(t+\Delta t)\rangle=\sqrt{2\gamma_{eb}}\int d^3
r\,C_2(\rv)\,|\, \psi_{\rm sys}(t)\rangle\,.
\label{eq:meas2}
\eeq
After each detection the process starts again from the beginning.

The conditional probability for the detection of the $(n+1)^{\rm th}$ photon
during the time interval $[t+t_{n},t+t_{n}+\delta t]$, given that the $n^{\rm th}$
detection occured at the time $t_{n}$, follows from Eqs.~{(\ref{eq:nonher})},
{(\ref{eq:prob1})}, and {(\ref{eq:prob2})}
\beq
P(t+t_{n};t_{n})=P_-(t+t_{n};t_n)+P_+(t+t_{n};t_n)\,,
\label{eq:totalprob}
\eeq
where the expression for $P_-(t+t_{n};t_n)$ is given explicitly by
\begin{mathletters}
\bea 
\lefteqn{P_-(t+t_{n};t_n)}\nonumber\\ &&= 2\gamma_{eb}\int d^3 r\,{\langle
\psi_{\rm sys}(t_n)\, |\,\exp{\{i H^\dagger_{\rm eff}t/\hbar\}}\,
C^\dagger_1(\rv)C_1(\rv)\,\exp{\{-i H_{\rm eff}t/\hbar\}}\, |\, \psi_{\rm
sys}(t_n)\rangle \over  \langle \psi_{\rm sys}(t_n)\, |\,  
\exp{\{iH_{\rm eff}^\dagger t/\hbar \}}\exp{\{-i H_{\rm eff} t/\hbar \}}\,|\, 
\psi_{\rm sys}(t_n)\rangle }\,\delta t\label{eq:cprob1a}\\
&&\simeq 2\gamma_{eb}\int d^3 r\,\langle
\psi_{\rm sys}(t_n)\, |\,e^{-i \tilde{H}_S t_n/\hbar}\,
C^\dagger_1(\rv, t+t_n)C_1(\rv, t+t_n)\,e^{i \tilde{H}_S t_n/\hbar}\, |\,
\psi_{\rm sys}(t_n)\rangle\, \delta t\,.  
\label{eq:cprob1b}
\eea 
\label{eq:cprob1}
\end{mathletters}
Here we have written the operator $C_1(\rv t)\equiv \exp{\{i
\tilde{H}_{S}t/\hbar\}}C_1(\rv)\exp{\{-i \tilde{H}_{S}t/\hbar\}}$ in the 
Heisenberg picture with respect to the system Hamiltonian $\tilde{H}_S$, defined 
in Eq.~{(\ref{eq:nsys})}.  Equation (\ref{eq:cprob1b}) is valid, if the
characteristic time evolution of the system Hamiltonian is much faster than the
detection rate of the spontaneously emitted photons \cite{JAC96}. The expression 
(\ref{eq:cprob1b}) is especially useful, if the measurements of the
spontaneously scattered photons Eqs.~{(\ref{eq:meas1})} and {(\ref{eq:meas2})}
do not significantly disturb the system, {\it i.e}. if 
$$
e^{i \tilde{H}_S t_{n+1}/\hbar}\,
|\, \psi_{\rm sys}(t_{n+1})\rangle \simeq e^{i \tilde{H}_S t_n/\hbar}\, |\,
\psi_{\rm sys}(t_n)\rangle\,.
$$ 
Then, the dynamics of the conditional probability $P_-(t;t_n)$ for detecting a
photon with the polarizability $\sigma_-$ during the time interval $[t,t+\delta
t]$ after detecting the $n^{\rm th}$ photon at the time $t_n$ is determined by the
system evolution of $C^\dagger_1(\rv t) C_1(\rv t)$. This system evolution may be
solved analytically from Eq.~{(\ref{eq:osc})}. We assume that the expectation
values for the number operators in both condensates are initially equal
$N_b=N_c=N/2$, and that $N\gg1$. Then, the expectation value $\langle
C^\dagger_1(\rv t) C_1(\rv t)\rangle_{\rm SB}$ in the spontaneous symmetry
breaking state, where the condensate annihilation operators have nonvanishing
expectation values Eq.~{(\ref{eq:SSB})}, and the expectation value $\langle
C^\dagger_1(\rv t) C_1(\rv t)\rangle_{\rm N}$ in the pure number state are given
by   \begin{mathletters}
\bea
\langle C^\dagger_1(\rv t) C_1(\rv t)\rangle_{\rm SB} &=& {N^2\over 16}
\left\{4 w(\rv)\, \left[1+\sin{(2\kappa t)}\sin{(\varphi)}\right]^2\right. 
\nonumber\\ && +v(\rv)\, [3+\cos{ (4\kappa t)}-\cos{ (4\kappa t)}
\cos{(2\varphi)} +\cos{(2\varphi)}] \nonumber\\
&&\left.+8 s(\rv)\, [1+\sin{ (2\kappa t)}\sin{ (\varphi)}]
\cos{( \varphi)} \right\}\,,
\label{eq:probosc} \\
\langle C^\dagger_1(\rv t) C_1(\rv t)\rangle_{\rm N} &\simeq &{N^2\over 16}
\left\{2 w(\rv)\, \left[2+\sin^2{(2\kappa t)}\right] 
 +v(\rv)\, [3+\cos{ (4\kappa t)} ]  \right\}\,.
\label{eq:proboscn} 
\eea
\label{eq:kprobosc} 
\end{mathletters}
Here we have used the following definitions
\beq
w(\rv)\equiv|\phi(\rv)|^4{\delta_1\over\Delta_1},\quad
v(\rv)\equiv |\phi(\rv)|^4 {\delta_2\over\Delta_1},\quad
s(\rv)\equiv {{\cal E}^*_1{\cal E}_2d_{eb}d_{ec}\over
4\hbar^2\Delta_1^2}|\phi(\rv)|^4  e^{-i {\bbox
\kappa}_{12} \cdot{\bf r}} \,.
\label{eq:probrel}
\eeq
To simplify the expressions, we have set in Eqs.~{(\ref{eq:kprobosc})} and
{(\ref{eq:probrel})} $\bar{\delta}=0$, $\phi_b(\rv)=\phi_c(\rv)\equiv\phi(\rv)$,
and $s(\rv)$ real. We can simplify Eq.~{(\ref{eq:kprobosc})} further by setting
$w(\rv)=v(\rv)=s(\rv)$  
\begin{mathletters}
\bea
\langle C^\dagger_1(\rv t) C_1(\rv t)\rangle_{\rm SB} &=& N^2 w(\rv)\,
\cos^2{\left({\varphi\over 2}\right)}\,[1+\sin{(2\kappa t)}\sin{(\varphi)}]\,,  
\label{eq:aveprobosc2} \\
\langle C^\dagger_1(\rv t) C_1(\rv t)\rangle_{\rm N} &\simeq &
{N^2 w(\rv)\over 2}\,.
\label{eq:aveproboscn2} 
\eea
\label{eq:aveprob2}
\end{mathletters}

The effects of macroscopic quantum coherence in Eqs.~{(\ref{eq:probosc})} and
{(\ref{eq:aveprobosc2})} are clearly observed, as they are in the Josephson
coupling, in Eq.~{(\ref{eq:jose})}. If the relative phase between the two
condensates $\varphi=\pi$, the expectation value in Eq.~{(\ref{eq:aveprobosc2})}
completely vanishes for the perfect spatial overlap of the condensate wave
functions. Then, the conditional probability $P_-(t;t_n)$ in 
Eq.~{(\ref{eq:cprob1b})} for detecting a photon with the polarizability
$\sigma_-$ during the time interval $[t,t+\delta t]$ after detecting the $n^{\rm
th}$ photon at the time $t_n$ also vanishes. If $\varphi=0$, the expectation value
{(\ref{eq:aveprobosc2})} is time-independent and twice as much as the
expectation value {(\ref{eq:aveproboscn2})} for the number state. For simplicity,
we set $\bar{\delta}=0$ in Eqs.~{(\ref{eq:kprobosc})}, {(\ref{eq:probrel})}, and
{(\ref{eq:aveprob2})}. These expressions are valid provided that the frequency of
the detections is much larger than $\bar{\delta}$. 
If the detuning from the two photon resonance becomes small, the effective
linewidth $\bar{\gamma}$ of the transition $c\rightarrow b$ may have an effect.
However, it may be shown to be proportional to $\Delta_1^{-2}$ or smaller.
The same expressions for $C^\dagger_2 C_2$ may be obtained from
Eq.~{(\ref{eq:aveprob2})} by changing $\varphi\rightarrow -\varphi$. 

Because in the scattering processes of atoms between the condensates the initial
states are indistinguishable, {\it i.e.} we do not know which condensate the
atoms are scattered from, the transition amplitudes interfere. For a large number
of atoms these interference effects may only be observed, if there is a
macroscopic quantum coherence of atoms present in the initial states, as in
Eqs.~{(\ref{eq:probosc})} and {(\ref{eq:aveprobosc2})}. In Ref.~{\cite{RUO97b}}
we investigated incoherent scattering, in which case there is an uncertainty in
both the initial and the final states of atoms. The macroscopic quantum
interference of the transition amplitudes affects both the scatterings of
atoms into the condensates and out of the condensates.

In this section we have shown that the macroscopic quantum coherence has
clearly observable effects on the conditional probability $P(t;t_n)$ in 
Eq.~{(\ref{eq:totalprob})} for spontaneous emission of photons. What remains to
be shown is that the measurement process of the spontaneously scattered photons
duly reaches a steady-state in which it does not significantly disturb the
dynamics of the system, and that this steady-state coincides with the theory of
the spontaneous breakdown of the gauge symmetry in BEC.

\section{Results of simulations}
\label{res}

We have simulated the light matter dynamics using the algorithms described in
Sec.~{\ref{stoch}}. The two condensates are assumed to be initially in the
number states with equal atom numbers, $N_b=N_c=N/2$. We define the visibility
of the macroscopic quantum interference $\beta$ and the angle $\varphi$ as the
modulus and the phase of the complex number after $n$ detections \beq \beta
e^{i\varphi}\equiv {2\over N}\langle\psi_{\rm sys}(t_n)\, |\,e^{-i \tilde{H}_S
t_n/\hbar}\, b^\dagger(0)c(0)\,e^{i \tilde{H}_S t_n/\hbar}\,|\, \psi_{\rm
sys}(t_n)\rangle\,. 
\eeq
According to the spontaneous symmetry breaking
arguments Eq.~{(\ref{eq:SSB})}, for a coherent state we have $\beta=1$, and
$\varphi=\varphi_c-\varphi_b$ is the relative macroscopic phase between the two
condensates. However, for number states there is no phase information at all
and $\beta=0$ before any detections are made. Starting from pure number states the
simulations show that a macroscopic quantum coherence is established by
measurements of spontaneously scattered photons. In each run of measurements the
relative phase between the condensates is selected as a random number. Although
the symmetry is broken in each individual run of measurements, it is regained
when an ensemble of measurements is considered.

With very small values of the effective two-photon detuning the system approaches
a dark state and the time between the photon detections increases rapidly.
In the dark state the driving light does not excite the atoms. The coherent
quantum tunneling between the condensates Eq.~{(\ref{eq:jose})} dies out and
the conditional probability of detecting spontaneously emitted photons,
Eqs.~{(\ref{eq:cprob1b})} and {(\ref{eq:aveprobosc2})}, goes to zero. This
corresponds to a vanishing relative phase between the condensates. On the
other hand, if the effective two-photon detuning is very large, the measurements
drive the system into a state where most of the particles are in one of the two
condensates. The maximum visibility of the interference between the condensates is
reduced from one to $\beta_{\rm max}=2\sqrt{N_b N_c}/N$, if the condensates have
unequal atom numbers. According to Bose-Einstein statistics, the scattering
to an already occupied state is enhanced. If most of the atoms are in the
condensate $b$, the probability $P_-$ of detecting a photon with the polarization
$\sigma_-$ Eq.~{(\ref{eq:prob1})} is much larger than $P_+$, the probability of
detecting a photon with the polarization $\sigma_+$ Eq.~{(\ref{eq:prob2})}. For
$N_b\gg N_c$, we also have $\langle C_1^\dagger C_1\rangle \sim \langle b^\dagger
b b^\dagger b\rangle$. Thus, the measurements end up driving the system towards a
number state reducing the macroscopic coherence. However, even with these two
extreme values of the effective two-photon detuning the measurements build up a
macroscopic quantum coherence with a randomly-defined relative phase between the
two condensates, before the coherence starts decreasing or the original phase is
lost. This is an example of the similiar behaviour of an initial coherent state
and an initial number state after a large number of measurements. The detections
rapidly build up a coherence and a random phase value for the initial number
state, even though these are not preserved for an arbitrary number of
measurements. In the absence  of collisions and with a carefully chosen
two-photon detuning the visibility $\beta\simeq 1$ may be maintained in the
simulations practically for an arbitrary number of detections.

We have run the simulations for the total number of 1000 atoms with the various
values of two-photon detuning $\delta_{cb}$. The light-induced level shifts
$\delta_1$ and $\delta_2$ from Eq.~{(\ref{eq:para})} are chosen equal to the
coupling coefficient $\kappa$, which is approximately $5\times 10^6$ times the
conditional probability of detecting the scattered photons from the pure number
state $N_b=N_c$ in Eq.~{(\ref{eq:totalprob})}. We have assumed that the external
wave functions of the condensates are equal $\phi_b(\rv)=\phi_c(\rv)$. 

In Fig. \ref{simfig1} we have shown the results of simulations of one run of 1000
detections. The two-photon detuning is given by $\delta_{cb}=0.15\kappa$. The
visibility $\beta$ in Fig. \ref{simfig1}(a) approaches to one very rapidly as the
number of detections is increased. The small decrease in the visibility after
700 detections is a consequence of unequal atom numbers in the condensates, as
the maximum visibility is reduced from one to $\beta_{\rm max}$. In Fig.
\ref{simfig1}(b) $\beta\sin\varphi$ also becomes well-defined as more detections
are made.

In Fig. \ref{simfig2} we have an example where the two-level detuning is larger,
$\delta_{cb}=2.0\kappa$. The visibility $\beta$ in Fig. \ref{simfig2}(a) is close
to one after 100 detections, but it starts then slowly decreasing as more atoms
enter the Zeeman state $b$. The maximum visibility of the interference between
the condensates is reduced, because of the unequal atom numbers. In Fig.
\ref{simfig2}(b) we have plotted the relative visibility defined by $\beta_r\equiv
\beta/\beta_{\rm max}$. This remains close to one. The relative phase between the
condensates in Fig. \ref{simfig2}(c) is reasonably steady.

The results of simulations for the two-level detuning $\delta_{cb}=0.05\kappa$
are shown in Fig.~{\ref{simfig3}}. The system approaches a dark state due to
spontaneous emission. The time elapsed from the beginning of the simulation
increases in Fig.~{\ref{simfig3}}(a) very rapidly as a function of the number of
measurements. According to Eqs.~{(\ref{eq:totalprob})}, {(\ref{eq:probosc})}, and
{(\ref{eq:aveprobosc2})} the vanishing expectation value for the conditional
probability of observing photons corresponds to a vanishing relative phase
between the condensates. The value of the phase $\varphi$ in
Fig.~{\ref{simfig3}}(b) rapidly approaches zero as the detection time starts
increasing, although the visibility $\beta$ remains close to one in
Fig.~{\ref{simfig3}}(c). This is an example of the strong dependence of the state
of the system on the  measurement scheme. Although the measurements first define
a random phase, the final value of the phase is not a random number. The
detections of the spontaneously emitted photons drive the system into a state in
which the relative phase between the two condensates vanishes. 

The conditional probability for the detection of the  $(n+1)^{\rm
th}$ photon during the time interval $[t,t+\delta t]$, given that the $n^{\rm th}$
detection occured at the time $t_{n}$, Eq.~{(\ref{eq:totalprob})}, does not have a
dependence on the relative phase of the two condensates, as in
Eqs.~{(\ref{eq:probosc})} and {(\ref{eq:aveprobosc2})}, before the macroscopic
quantum coherence is established by measurements. According to the conventional
symmetry breaking arguments such a phase dependence should exist {\it immediately
from the first measurement}. Given that the time between photon detections could
be determined very accurately, a different behaviour would be observed, whether
the condensates are initially in  number states or coherent states.
Similarly, if we start from pure number states, there is no coherent quantum
tunneling between the condensates according to  Eq.~{(\ref{eq:jose})} before
$\beta\neq 0$.

According to Eq.~{(\ref{eq:aveprob2})} the conditional probability $P(t;t_n)$ 
for detecting spontaneously emitted photons can completely vanish for the
spontaneous symmetry breaking state, while it is always finite for a number state.
If the relative phase between the condensates is equal to zero, no photons in the
spontaneous symmetry breaking state are observed for perfect spatial overlap
of the condensate wave functions. However, if the condensates were initially in
number states, emitted photons would be detected according to  
Eq.~{(\ref{eq:aveproboscn2})} before the macroscopic coherence is established by
measurements, even if this coherence suppressed the emission of photons. In our
simulations for a thousand atoms the phase becomes well-defined after
approximately 100-200 detections. With the total number of atoms $N\leq1000$, the
required number of measurements seems to be well above $\sqrt{N}$. For 
condensates formed with from $10^6$ to $10^8$ atoms, a significantly different
behaviour is observed, whether the condensates are initially in number states or
in coherent states. This is an alternative scheme for testing Bose broken
symmetry arguments to that suggested by Wong {\it et al.} \cite{WON97}. They
proposed that the validity of the Bose-broken symmetry could be tested from the
collapses and revivals of the macroscopic quantum coherence \cite{WRI96,WRI97},
if atoms are not lost from the system during the measurement process.

For our light scattering scheme to be a practical test of the validity of the
spontaneous symmetry breaking arguments for dilute atomic gases, the spatial
overlap of the condensate wavefunctions should be significant and the phase
diffusion by collisions between different atoms should be slow compared to the
spontaneous emission rate. According to Eq.~{(\ref{eq:aveprob2})} the probability
of detecting spontaneously emitted photons as a function of time is stimulated by
the large number of atoms in the condensates. Depending on the geometry of the
system the spontaneous emission rate is roughly proportional to $\gamma N_e
N\simeq\gamma N^2\delta_1/\Delta_1$, for $\delta_1 \sim\delta_2$. Here $N$ is the
total number of atoms in the condensates, $\gamma$ is the free space spontaneous
emission rate, $N_e$ is the number of electronically excited atoms, and
$\delta_1$ and $\delta_2$ are the light induced level-shifts from
Eq.~{(\ref{eq:para})}. The detuning of the incoming light $\Delta_1$ was chosen
large. Wong {\it et al.} \cite{WON96} have studied the effect of collisions on
the relative phase between two Bose condensates. The two-body collision rate can
be estimated by $\kappa\sim\rho\pi a^2 v_{\rm rms}$, where $a$ is the scattering
length, $\rho$ is the density of atoms, and $v_{\rm rms}$ is the root-mean-square
speed of atoms. At a temperature of 180 nK and with a density of $10^{12}$
cm$^{-3}$ this gives for $^{87}$Rb atoms a collision rate of about one
collision per second. In Ref.~{\cite{WON96}} the simulations for 200 atoms show a
recognizable steady conditional phase for the collision rate
$\kappa=0.2\Gamma_a$, where $\Gamma_a$ is the atom detection rate. A strong
overlap between the condensates could significantly reduce the decoherence
effects of the collisions.

\section{Conclusions}
\label{conc}

We have studied carefully the interaction of light with two Bose condensates as an
open quantum system including the effects of measurements. In the limit of large
detuning of the driving light beams from the atomic resonances, we have shown
that the relative phase between the two Bose condensates may be established by
the measurements of spontaneously scattered photons, even though the condensates
are initially in pure number states. In the quantum trajectory simulations the
strong effect of measurements on the state of the condensates is clearly
observed. Even the modification of the effective two-photon detuning changes
substantially the effects of the measurements.

The measurement of spontaneously scattered photons is non-destructive for the
condensates, because only light is scattered, but atoms are not removed from the
two condensates. In particular, a non-destructive detection allows repetitions of
independent runs of measurements for the same condensates at different times. 
Establishing the relative phase between the condensates by the simulations of
measurements on scattered light has also other advantages over the simulations of
atom counting. In the case of light scattering we can use the well-known theories
of photon detection \cite{GLA63}. The present authors are not aware of the
existence of similiar theories for atom detection. The driving light also
introduces naturally the high frequencies for the system dynamics, which are
required for the Markov and Born approximations in the derivation of the
stochastic Schr\"odinger equations. Thus, we have obtained the first evidence of
the macroscopic coherence properties of Bose condensates, initially in number
states, by simulations which may be justified by well-established physical
theories. Although the simulations of atom counting have previously shown
coherence properties, it was not evident {\it a priori} that the photon
detections also should, because the different measurements may affect the system
in a very different way.

We have shown that the conditional probability of detecting spontaneously
scattered photons as a function of time depends strongly on the relative phase
between the two Bose condensates. This may provide a method to detect the relative
phase and to give an unambigious signature of the macroscopic quantum coherence
in BEC. The significant dependence of the spontaneous emission rate on the
condensate phase difference could possibly also be used as a test of the
spontaneous symmetry breaking of the global gauge invariance in dilute atomic
gases, {\it i.e.} as a way to determine the true quantum state of the Bose
condensate.

\subsection*{Acknowledgements}
We would like to thank Matthew Collett and Michael Jack for usuful discussions.
This work was supported by the Marsden Fund of the Royal Society of
New Zealand, The University of Auckland Research Fund and The New Zealand
Lottery Grants Board.

\begin{figure}
\caption{
The level scheme of the system. Two Bose condensates are in two different Zeeman
sublevels $|b\rangle=|g,m\rangle$ and $|c\rangle=|g,m-2\rangle$. The states 
$|c\rangle$ and $|b\rangle$ are optically coupled to the
center-of-mass manifold of the electronically excited state
$|e\rangle=|e,m-1\rangle$ by the far-off-resonant driving fields $\Ev_{d2}$ and
$\Ev_{d1}$ having polarizations $\sigma_+$ and $\sigma_-$, respectively. } 
\label{levelfig}    
\end{figure}

\begin{figure}
\caption{
One run of simulations for 1000 atoms. (a) shows how the visibility of the
macroscopic quantum interference $\beta$ approaches one as a function of the
number of measured photons. In (b) we have plotted $\beta\sin\varphi$ as a
function of the number of detections. The relative phase between the two Bose
condensates $\varphi$ also becomes well-defined. Here we have set
$\delta_1=\delta_2=\kappa$ and the two-level detuning  $\delta_{cb}=0.15\kappa$.
} 
\label{simfig1}   
\end{figure}

\begin{figure}
\caption{
One run of simulations for 1000 atoms with $\delta_1=\delta_2=\kappa$ and the
two-level detuning  $\delta_{cb}=2.0\kappa$. In (a) the visibility $\beta$ is
close to one after 100 detections of spontaneously scattered photons, but starts
then decreasing, because of the unequal number of atoms in the two condensates.
However, the relative visibility $\beta_r$ in (b) remains close to one. The
relative phase between the two condensates is plotted in (c).
 }  
\label{simfig2} 
\end{figure}

\begin{figure}
\caption{
One run of simulations for 1000 atoms with $\delta_1=\delta_2=\kappa$ and the
two-level detuning  $\delta_{cb}=0.05\kappa$. The time elapsed from the
beginning of the simulation as a function of the detected photons in (a) starts
increasing as the system approaches a dark state. The value of the phase
$\varphi$ in (b) approaches rapidly zero as the detection time starts increasing,
although the visibility $\beta$ in (c) remains close to one.
 }  
\label{simfig3} 
\end{figure}

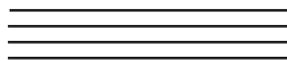
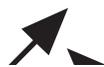
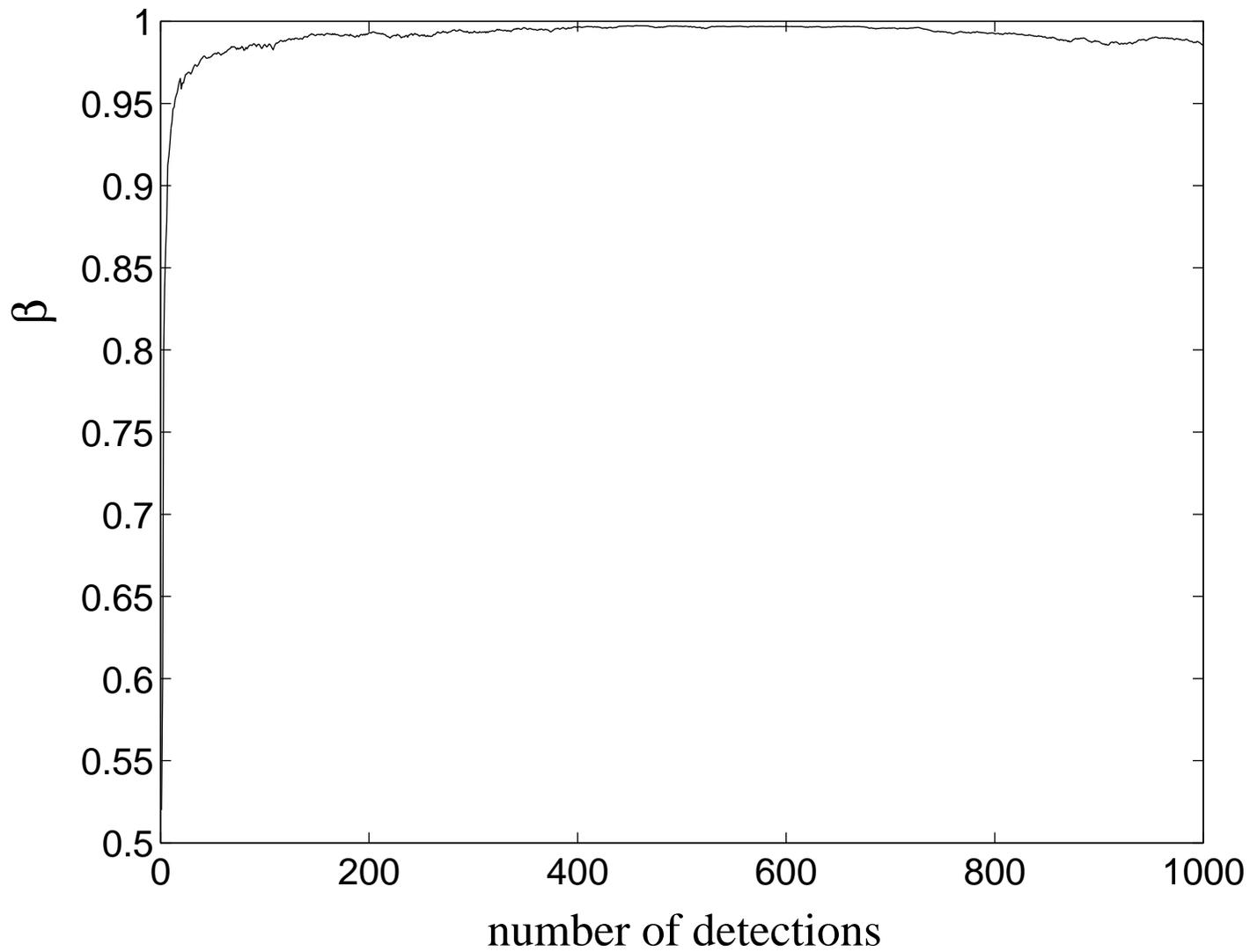
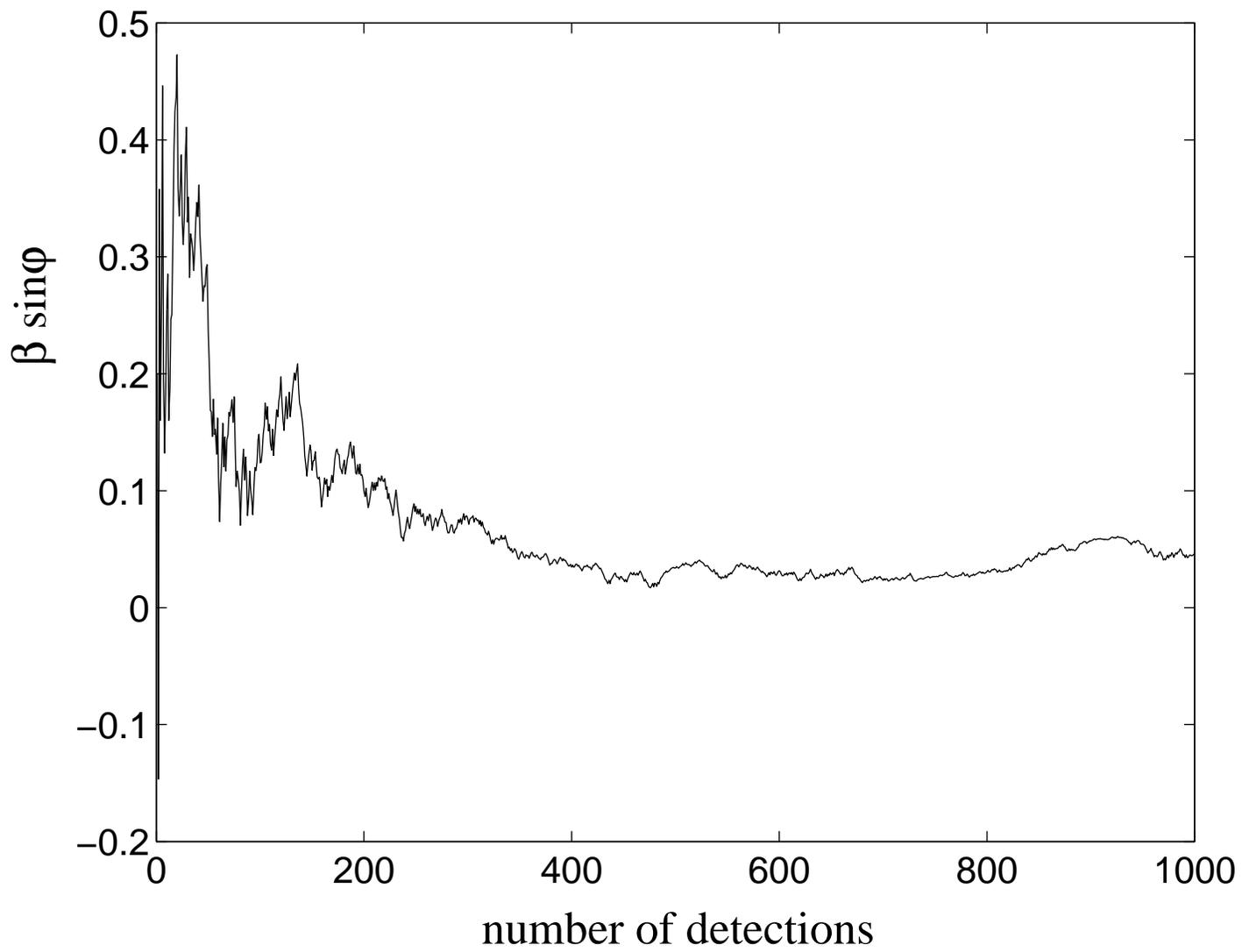
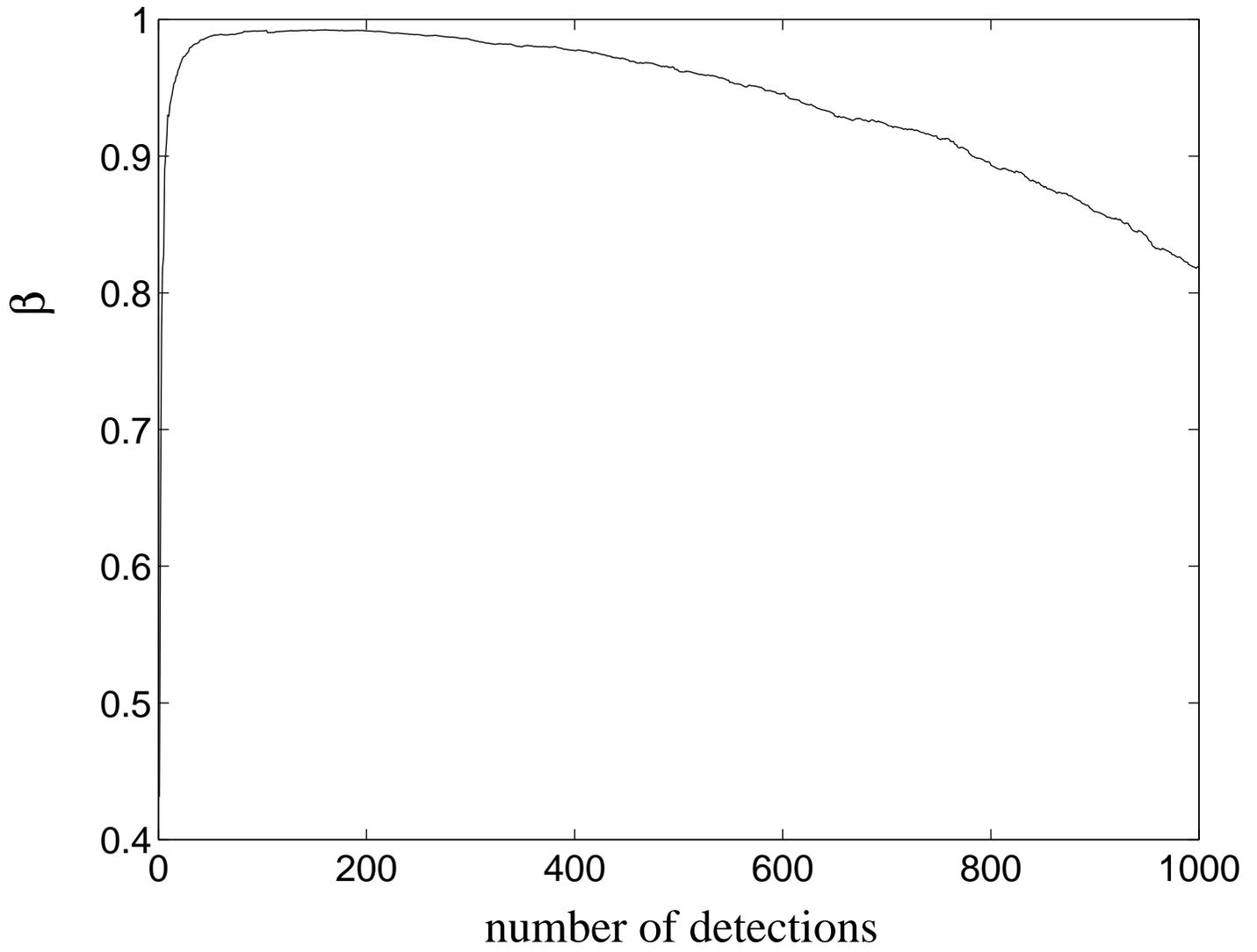
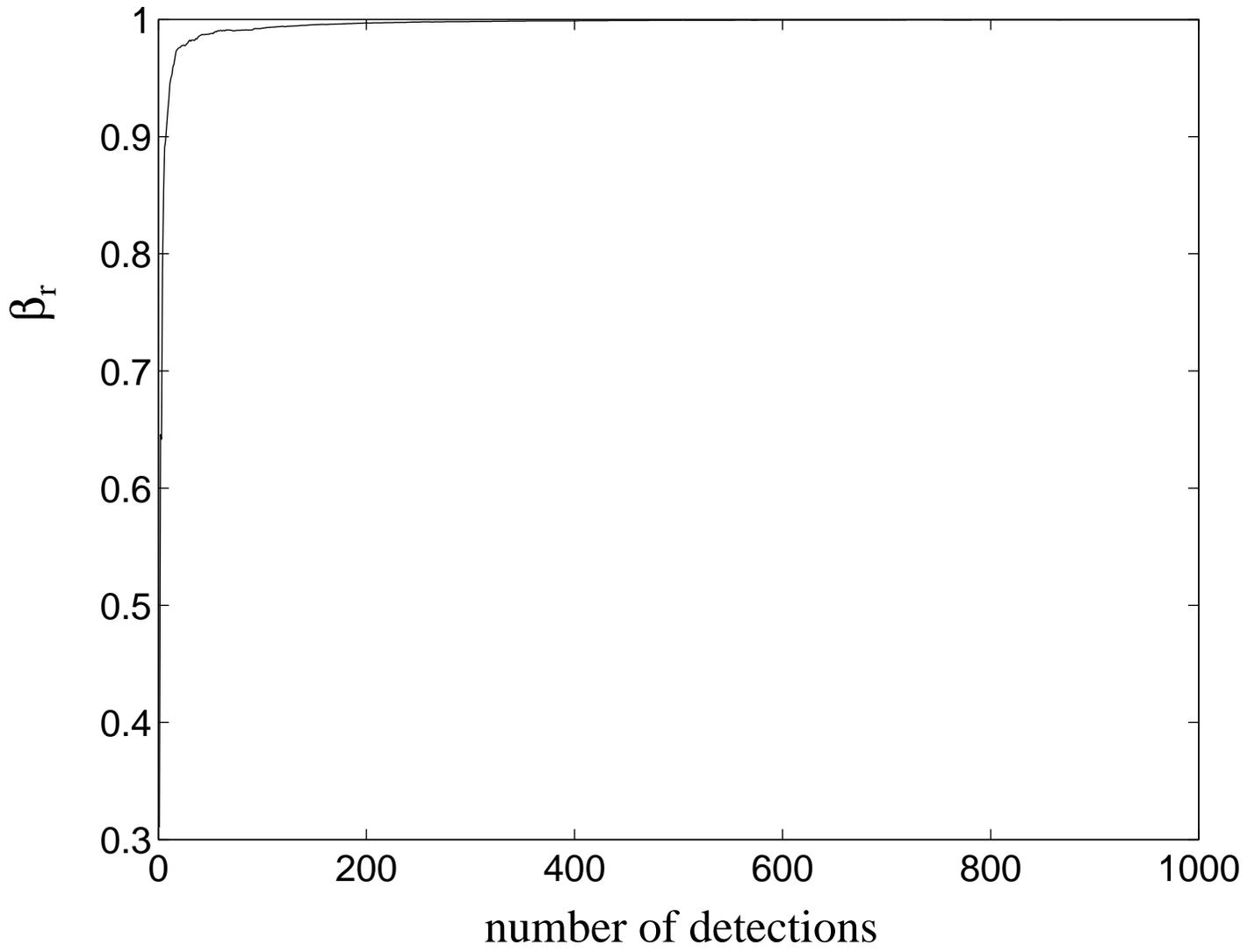
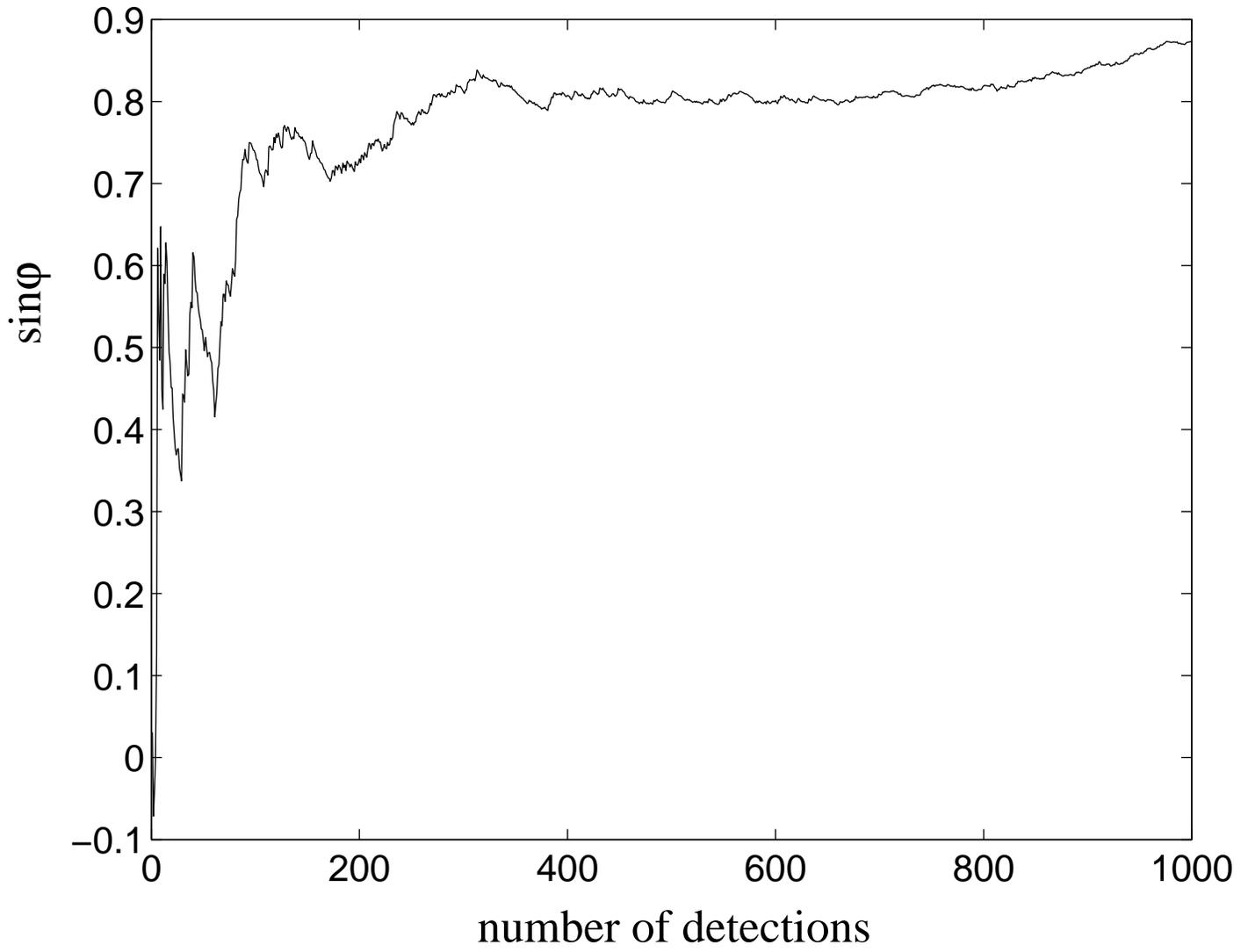
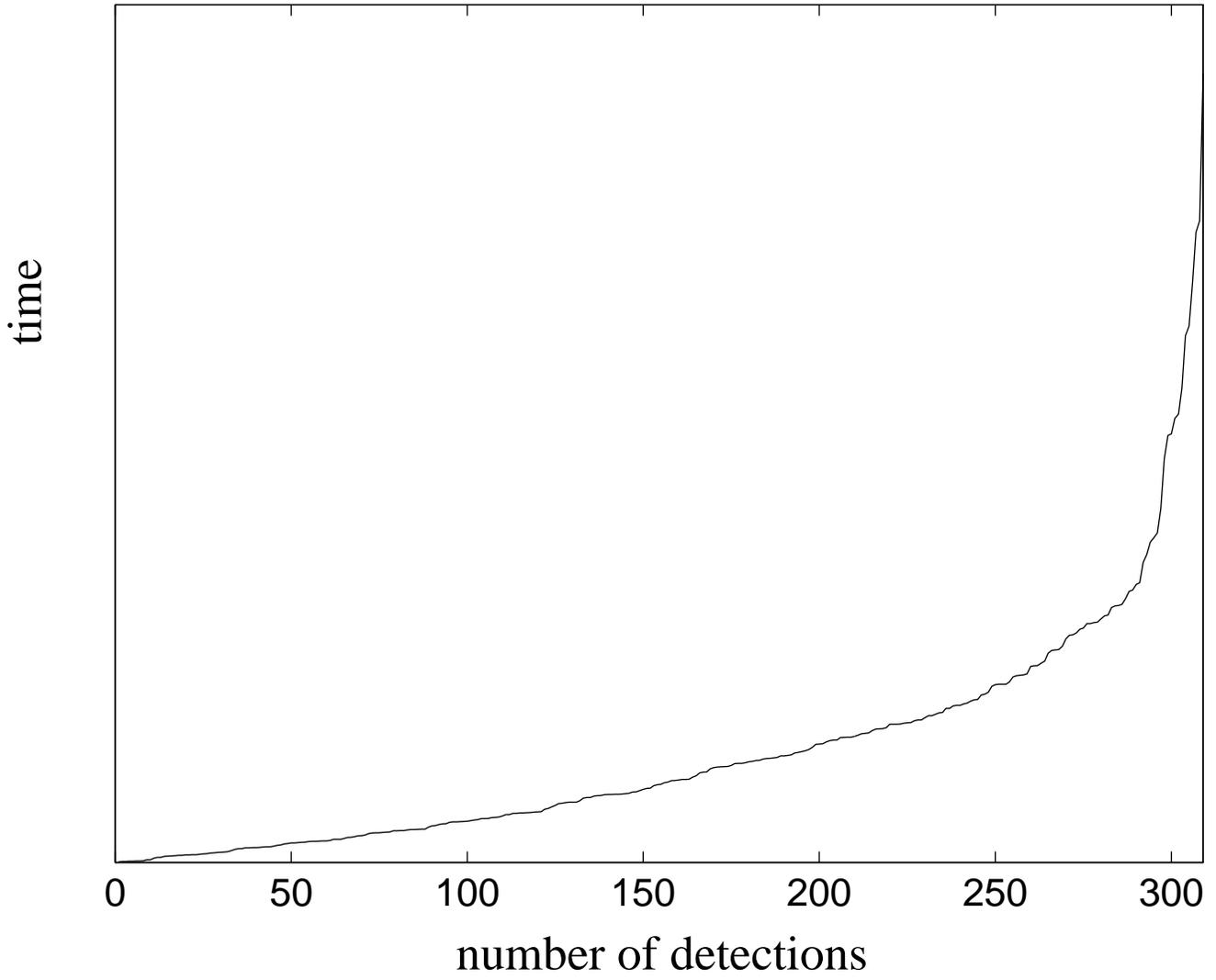
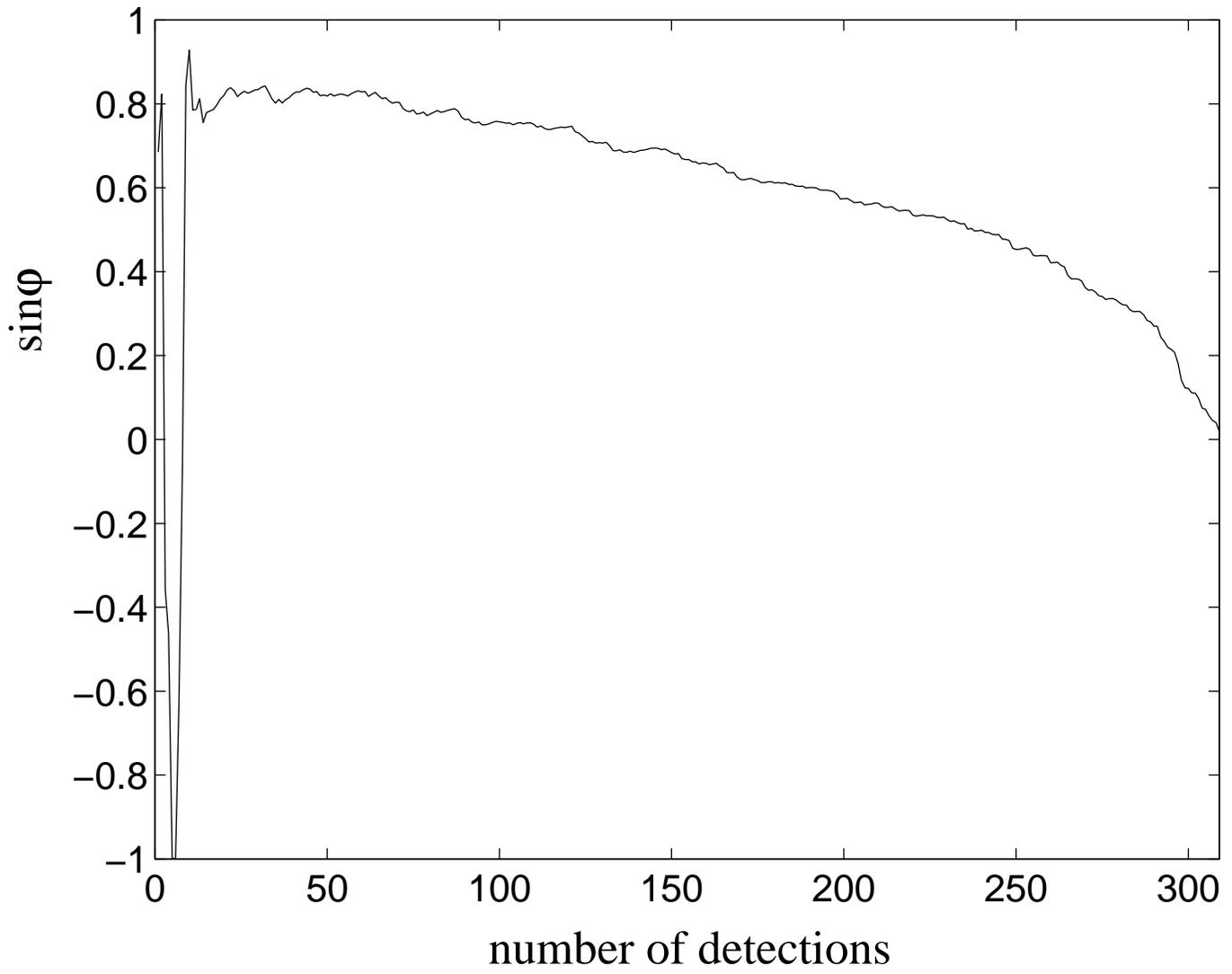
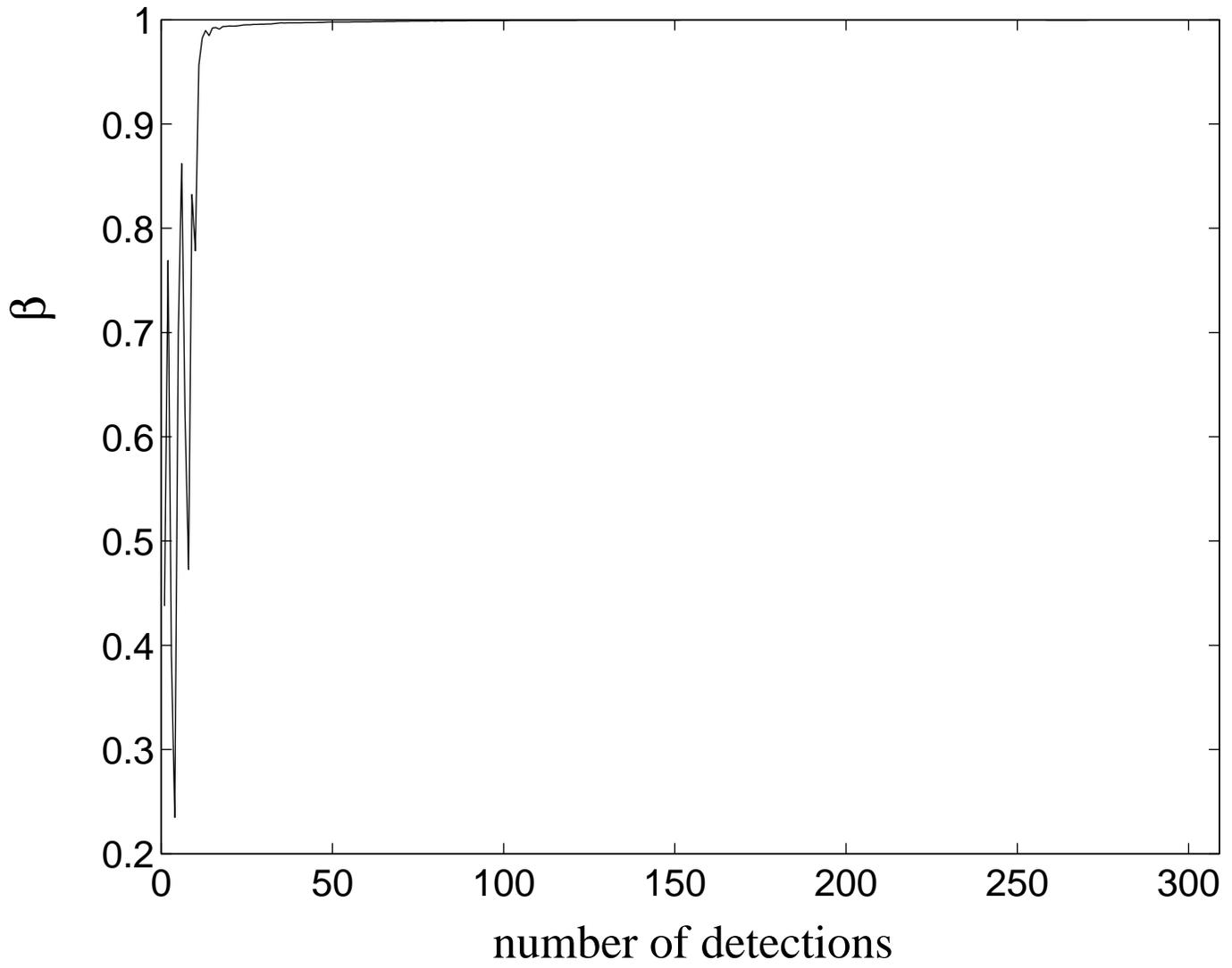
\end{document}